\newif\ifTwoColumn
\newif\ifSUBMIT
\newif\ifCOMMENTS
\newif\ifFIGs
\newif\ifFIGoneColumn
\let\ifTwoColumn\iftrue
\let\ifSUBMIT\iftrue
\let\ifCOMMENTS\iffalse
\let\ifFIGs\iftrue
\let\ifFIGoneColumn\iftrue
\ifTwoColumn
  \documentclass[aps,showpacs,prb,showkeys,twocolumn,superscriptaddress,
     float,floatfix]{revtex4-1} %'prb' removed -> citation superscript
\else
  \documentclass{elsarticle}
\fi
\usepackage{graphicx}
\usepackage{amsmath}
\usepackage{xcolor}
\usepackage{hyperref}
\usepackage[utf8]{inputenc}

\usepackage{array,mathtools,amssymb,booktabs}
\ifSUBMIT
  \ifCOMMENTS
    \usepackage{color}    %%% NOTE this is necessary in pdflatex
    \usepackage[normalem]{ulem}
    \def\EDITS#1{{\color{green}#1}}
    \def\STRIKE#1{{\color{red}\sout{#1}}}
    
    \def\NSTRIKE#1{{\color{blue}\sout{#1}}}
  \else
    \def\EDITS#1{#1}
    \def\STRIKE#1{}
    
    \def\NSTRIKE#1{}
  \fi
\else
  \usepackage{color}    %%% NOTE this is necessary in pdflatex
  \usepackage[normalem]{ulem}
 \definecolor{mygreen}{RGB}{0,180,0}    %%% NOTE that this isn't necessary.
  \def\EDITS#1{{\color{mygreen}#1}}
  \def\STRIKE#1{{\color{red}\sout{#1}}}

  \def\NSTRIKE#1{{\color{blue}\sout{#1}}}
\fi

\definecolor{mygray}{RGB}{128,128,128}

\DeclareMathOperator{\argmin}{arg\,min}

\begin{document}

\title{Variational principle for the determination of unstable periodic orbits 
and instanton trajectories at saddle points}
\author{Andrej Junginger}
\author{Jörg Main}
\author{Günter Wunner}
\affiliation{%
Institut f\"ur Theoretische Physik 1,
Universit\"at Stuttgart, 
70550 Stuttgart,
Germany}

\author{Rigoberto Hernandez}
\email [Correspondence to: Rigoberto Hernandez,
Department of Chemistry, Johns Hopkins University,
Baltimore, MD 21218
E-mail: ]{r.hernandez@jhu.edu}
\affiliation{%
Department of Chemistry,
The Johns Hopkins University,
Baltimore, MD 
}

\date{\today}

%%% user=defined commands %%%%%%%%%%%%%%%%%%%%%%%%%%%%%%%%%%%%%%%%%%%%%%%%%%%% 
\newcommand{\EQ}{Eq.}
\newcommand{\EQS}{Eqs.}
\newcommand{\FIG}{Fig.}
\newcommand{\FIGS}{Figs.}
\newcommand{\REF}{Ref.}
\newcommand{\REFS}{Refs.}
\newcommand{\SEC}{Sec.}
\newcommand{\SECS}{Secs.}
\newcommand{\eg}{e.\,g.}
\newcommand{\cf}{cf.}
\newcommand{\ie}{i.\,e.}
\newcommand{\ud}{\mathrm{d}}
\newcommand{\ue}{\mathrm{e}}
\newcommand{\kB}{k_\mathrm{B}}
\newcommand{\VLiCN}{V_\mathrm{LiCN}}
\newcommand{\VCN}{V_\mathrm{C-N}}
\newcommand{\VLi}{V_\mathrm{Li-CN}}
\renewcommand{\vec}[1]{\boldsymbol{#1}}
\newcommand{\qq}{\vec{q}}
\newcommand{\xx}{\vec{x}}
\newcommand{\vv}{\vec{v}}
\newcommand{\transpose}{\mathsf{T}}
\newcommand{\reactantpop}{\mathcal{P}}
\newcommand{\kf}{k_\mathrm{f}}
\newcommand{\etal}{\emph{et al.}}
\newcommand{\LD}{\mathcal{L}}
\newcommand{\LDf}{\LD^\text{(f)}}
\newcommand{\LDb}{\LD^\text{(b)}}
\newcommand{\LDfb}{\LD^\text{(fb)}}
\newcommand{\LDfbw}{\LD^\text{(fbw)}}
\newcommand{\Ws}{\mathcal{W}_\text{s}}
\newcommand{\Wu}{\mathcal{W}_\text{u}}
\newcommand{\Wsu}{\mathcal{W}_\text{s,u}}
\newcommand{\TSt}{\mathcal{T}}
\newcommand{\PO}{\mathcal{P}}
\newcommand{\weightingf}{\chi^\text{(f)}}
\newcommand{\weightingb}{\chi^\text{(b)}}
\newcommand{\weightingfb}{\chi^\text{(f,b)}}
\newcommand{\vtherm}{v_\text{therm}}
\newcommand{\Esaddle}{E^\ddagger}
%%%%%%%%%%%%%%%%%%%%%%%%%%%%%%%%%%%%%%%%%%%%%%%%%%%%%%%%%%%%%%%%%%%%%%%%%%%%%%
\newcommand{\comment}[1]{\textsf{\textcolor{orange}{[#1]}}}
\newcommand{\sno}[1]{_\mathrm{#1}}
\newcommand{\no}[1]{\mathrm{#1}}
\newcommand{\acnew}[1]{\acfi{#1}\acused{#1}}

\begin{abstract}
The complexity of arbitray dynamical systems and 
chemical reactions, in particular, can often be resolved if only 
the appropriate periodic orbit---in the form of a 
limit cycle, dividing surface, instanton trajectories or some other 
related structure---can be uncovered.
Determining such a periodic orbit, no matter how beguilingly simple it appears,
is often very challenging.
We present a method for the direct construction of unstable periodic orbits and 
instanton trajectories at saddle points by means of Lagrangian descriptors.
Such structures result from the minimization of a scalar-valued 
phase space function without need for 
any additional constraints or knowledge.
We illustrate the approach for two-degree of freedom systems at a rank-1 saddle 
point of the underlying potential energy surface by constructing both periodic 
orbits at energies above the saddle point as well as instanton trajectories 
below the saddle point energy.
\end{abstract}

\maketitle

%%%%%%%%%%%%%%%%%%%%%%%%%%%%%%%%%%%%%%%%%%%%%%%%%%%%%%%%%%%%%%%%%%%%%%%%%%%%%%%%
\section{Introduction}

Periodic orbits (POs) 
are useful in resolving the overall dynamics 
in the disparate fields of 
chaotic dynamics,\cite{Stoeckmann1998,grob92,romek94,Ott2002a,Schuster2005,Uzer2006}
semiclassics,\cite{gutzwiller1970,gutzwiller1971,Gutzwiller1990,Brack96a}
classical reaction dynamics,\cite{pollak85,pech77,pollak78,pollak80a}
and tunneling processes.
\cite{miller75,cole77c,cole77a,affleck81,stoof97,Wunner08a,rommel11,
Einarsdottir2012,meisner2016}
A general distinction can be made between stable and unstable POs.
Integrable systems give rise to POs
whose oscillations on the invariant tori 
can be characterized by corresponding particular frequencies.
More generally, POs form limit cycles, 
are essential to Gutzwiller's trace formula\cite{gutzwiller1970,gutzwiller1971}
and are crucial in semiclassical quantization 
theories.\cite{mill74,mill77,hern93b,hern94}
Unstable POs are of importance in the field of classical reaction 
dynamics in classical systems
where they form recrossing-free dividing 
surfaces.\cite{pollak79,pollak85,grob92,romek94,dawn05a,hern10a}
They are important for understanding tunneling dynamics
through a barrier in quantum mechanical reactions
wherein the PO on the inverted potential, $-V$, is the
instanton trajectory providing the leading contribution to the
path integral.\cite{kleinert2009path}
In common between all of these cases, 
is the invariance of POs as they provide a scaffold from
which to obtain other geometric structures,
and thus remain objects of current interest such as
in, e.g., Refs.~\onlinecite{jancovic2016, cvitanovic2016}.

In many cases, the PO of interest is a rather simple trajectory, 
but a substantial effort is often needed to find it, if can be found at all.
The requirement of periodicity, $\xx(t) = \xx(t+T)$ with the characteristic 
period $T$, is usually employed as an appropriate constraint.
Specifically, root-search algorithms are employed to solve
$\xx(T) - \xx(0) = 0$
with respect to the boundary conditions:
the unknown initial position $\xx(0)$ on the PO,
and the unknown period $T$.
If the desired PO is a hyperbolic trajectory,
then numerical instabilities can derail the
convergence in the root-search procedure.
In some such cases, it may even be impossible to integrate the PO 
over the whole period within the numerical accuracy of 
standard integrators.
Multi-step integrators such as the 
advanced multi-shooting algorithm\cite{morrison1962,kiehl1994} 
can help to address some of these but not all.

In this paper, we present an alternate approach for determining
unstable POs at saddle points with energies above 
and below the corresponding saddle point energy $\Esaddle$.
We have found that such POs can be obtained 
through a variational procedure 
based on Lagrangian descriptors 
(LDs).\cite{Mancho2010,Mancho2013,hern14b,hern14f,hern15a,hern15e,hern16a,%
hern16h, hern16i}
The connection of the LD to the construction of POs is presented
in \SEC~\ref{sec:theory}.
An advantage of this procedure in avoiding the root-search algorithm
is that it requires no knowledge of the boundary conditions.
It is only based on minimizing the scalar-valued LD in phase space.
We demonstrate the LD-PO procedure in \SEC~\ref{sec:results}
through the determination of unstable POs
in two very different dynamical cases involving rank-1 saddle points.
The first example addresses classical POs at energies above the saddle 
point energy, $E>\Esaddle$, that are related to the dividing surface.
The second example constructs instanton trajectories 
below the saddle point energy, $E<\Esaddle$, that are 
related to quantum-mechanical tunneling.

\section{Construction of periodic orbits through Lagrangian descriptors} 
\label{sec:theory}

The LD is a scalar-valued function in phase space and for each point $( \vec 
x_0, \vec v_0)$ and time $t_0$. It is defined by
\begin{equation}
 \LD( \vec x_0, \vec v_0, t_0) = 
 \int_{t_0-\tau}^{t_0+\tau} \| \vec v(t) \| \, \mathrm{d}t \,.
 \label{eq:LD}
\end{equation}
Here, the integration time $\tau$ is a free parameter 
that has to be chosen large enough 
such that
underlying structures of the dynamics are sufficiently resolved,
and that all relevant time scales are covered.
For a given $\tau$, the resulting LD is a measure of the arc length of 
the respective trajectory $\qq(t)$ starting at $(\xx_0,\vv_0,t_0)$ in 
forward and backwards time.
In recent work,\cite{hern15e,hern16a,hern16d}
\EDITS{our group has demonstrated that 
the moving dividing surface associated with the transition
state trajectory is nonrecrossing.}
In arbitrary dimensions 
\EDITS{at each instance in time}, this structure is a normally hyperbolic 
invariant manifold (NHIM)\cite{Fenichel72,bristol2} which
can be obtained by perturbation 
theory.\cite{hern93b,hern94,Jaffe05,Komatsuzaki99,Waalkens04b,hern10a}
\EDITS{Using extremal values of the LD, 
we were also able to construct the DS, itself.\cite{hern15e,hern16a,hern16d}}
Meanwhile, 
\EDITS{at constant energy on two-dimensional stationary potentials,}
Pollak and Pechukas\cite{pollak78, pech79a,pollak80} 
showed that the nonrecrossing dividing surface
is a PO associated with \EDITS{the nearby} saddle point.
Thus the LD-PO conjecture arises from 
\EDITS{the logic of these connected arguments in two-dimensions:}
the trajectory which minimizes the LD
is a periodic orbit associated with a saddle point.

More generally, the essential connection between the LD and PO associated with
a saddle point
lies in
the fact that each orbit $\PO_E$ at energy $E$ is located at the intersection 
of the stable and unstable manifolds $\Wsu$ attached to the saddle,
\begin{equation}
  \PO_E = \Ws \cap \Wu \bigl.\bigr|_E \,.
  \label{eq:PO-intersec-manifolds}
\end{equation}
This relation holds because any deviation of the trajectory from 
the PO
with a nonzero unstable contribution would let the particle escape.
Moreover,
trajectories on both of these manifolds ($\Ws$ and $\Wu$) lead
to extremal values of the LD in \EQ~\eqref{eq:LD}.
More precisely, the manifolds correspond either to a minimum of the LD's 
forward 
($f$: $t_0\leq t \leq t_0+\tau$) 
or its backward 
($b$:  $t_0-\tau\leq t \leq t_0$) 
contribution,
\begin{subequations}
\begin{align}
 \Ws(t_0) &= \argmin \LDf (\vec x_0, \vec v_0,t_0) \,, \\
 \Wu(t_0) &= \argmin \LDb (\vec x_0, \vec v_0,t_0) \,,
\end{align}%
\label{eq:LD-Wsu}%
\end{subequations}
where `$\argmin$' denotes the argument of the minimum of the LD hypersurface 
with respect to the phase space coordinates ($\xx_0,\vv_0$). 
Thus the position of the PO in phase space is directly 
related to the minimum of the LD~\eqref{eq:LD}, leading
to the periodic orbit:
\begin{equation}
 \PO_E(t_0) = \argmin \LD (\vec x_0, \vec v_0, t_0) \bigl.\bigr|_E \,.
 \label{eq:LD-PO}
\end{equation}
In order to obtain the LD-PO, it is not necessary that the 
integration time $\tau$ of the LD coincides with the period $T$.
In a numerical scheme, it is therefore possible to start a PO search
optimization
with 
short integration times providing an approximation to the PO.
A subsequent increase of the integration times then improves the results and 
the numerical accuracy of the PO as needed.

\section{Applications to rank-1 saddles in two-degree of freedom systems} 
\label{sec:results}

We illustrate the LD-PO method
through two cases: the construction of the periodic orbit
dividing surface (PODS),\cite{pollak85,pech77,pollak78,pollak80a}
and the instanton trajectory.\cite{stoof97,Wunner08a}
All the following results have been obtained by numerically integrating the 
dynamical equations corresponding to the respective given potentials using a 
velocity verlet algorithm.
For the resulting trajectory, the LD \eqref{eq:LD} has been evaluated on the 
fly during the integration, and the minimization of the LD hypersurface in 
\EQ~\eqref{eq:LD-PO} has been performed using a standard simplex procedure.

\subsection{Determining the PODS in 2D}
\label{sec:results-classical}

We first apply the LD-PO method to construct 
unstable POs of a rank-1 saddle point in a two-degree of freedom system 
in the form of the model potential\cite{hern16T5}
\begin{align}
V(x,y) = 
\Esaddle\,\exp\left(-x^2\right) + 
\frac{\omega\sno{y}^2}{2}\left[y-\frac{2}{\pi}
\arctan\left(2 x \right)\right]^2
\label{eq:pot2d}
\end{align} 
with $\Esaddle=2$ and $\omega\sno{y}=2$ in arbitrary units.
This potential represents a Gaussian barrier in the $x$-direction that is 
nonlinearly coupled to a harmonic oscillator in the $y$-direction.
It can be 
seen as prototypical for a chemical reaction with reactants $x\to-\infty$ and 
products $x\to+\infty$ that are separated by an energy barrier.

%%%%%%%%%%%%%%%%%%%%%%%%%%%%%%%%%%%%%%%%%%%%%%%%%%%%%%%%%%%%%%%%%%%%%%%%%%%%%%%
\begin{figure}[t]
\centering
\includegraphics[width=0.9\columnwidth]{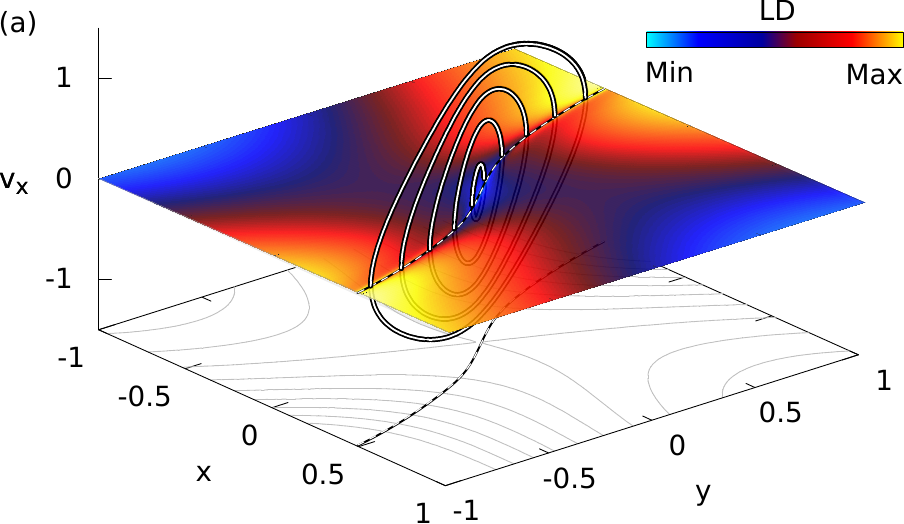}
\includegraphics[width=0.9\columnwidth]{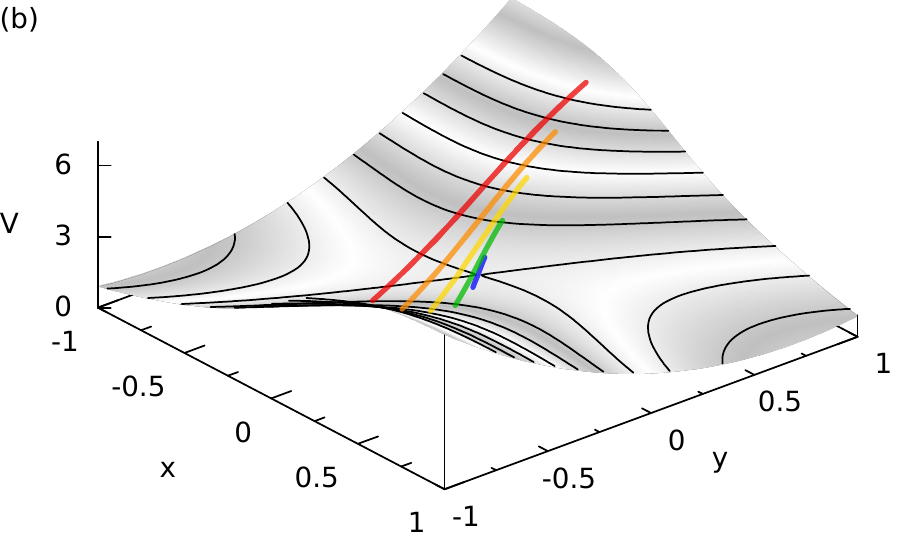}
\caption{%
Visualization of five POs (solid lines) at different energies above 
the activation energy.
Panel (a) shows an LD cut through phase space according to \EQ~\eqref{eq:LD} 
with $\tau=10$ and a selection of POs in $(x,y,v_x)$-space 
(analogous results are obtained also for $v_y$) 
and the LD is calculated for initial conditions $v_x=v_y=0$.
The LD exhibits a minimum valley (dashed black and white lines) whose phase 
space coordinates belong to the periodic orbit at the respective energy.
Panel (b) presents the configuration space projection of the POs together with 
the underlying potential.
The POs presented have periods in the range $2.19 \lesssim T \lesssim 2.45$, so 
that the integration time $\tau$ is sufficient to provide a high-resolution 
minimum LD valley.
}
\label{fig:PODS-energy}
\end{figure}
%%%%%%%%%%%%%%%%%%%%%%%%%%%%%%%%%%%%%%%%%%%%%%%%%%%%%%%%%%%%%%%%%%%%%%%%%%%%%%%

In \FIG~\ref{fig:PODS-energy}(a), we show a cut of the LD~\eqref{eq:LD} through 
the $x$-$y$-plane in phase space for an integration time $\tau=10$
as a colored contour plot.
(The selected value of $\tau$ is large compared to the periods $T$ 
of the POs shown in the figure which satisfy
$2.19 \lesssim T \lesssim 2.45$.)
For each energy $E$ there are two minima of the LD which correspond to the 
intersection of the PO with that plane. 
All the minima of the LD at different energies together form a valley which is 
highlighted by the dashed white line.
According to \EQ~\eqref{eq:LD-PO}, each point on this line is an initial 
condition of a PO at the respective energy.
Five POs selected at representative distinct energies $E>\Esaddle$ 
are shown in $x$-$y$-$v_x$-space with little loss of information 
because $v_y$ is not shown.
All five intersect with the $x$-$y$-plane at 
the LD's minimum valley as suggested by \EQ~\eqref{eq:LD-PO}.
Figure~\ref{fig:PODS-energy}(b) shows the configuration space projection of the 
POs together with the potential~\eqref{eq:pot2d} visualizing the location of 
the POs on the potential energy surface.

%%%%%%%%%%%%%%%%%%%%%%%%%%%%%%%%%%%%%%%%%%%%%%%%%%%%%%%%%%%%%%%%%%%%%%%%%%%%%%%
\begin{figure}[t]
\centering
\includegraphics[width=0.9\columnwidth]{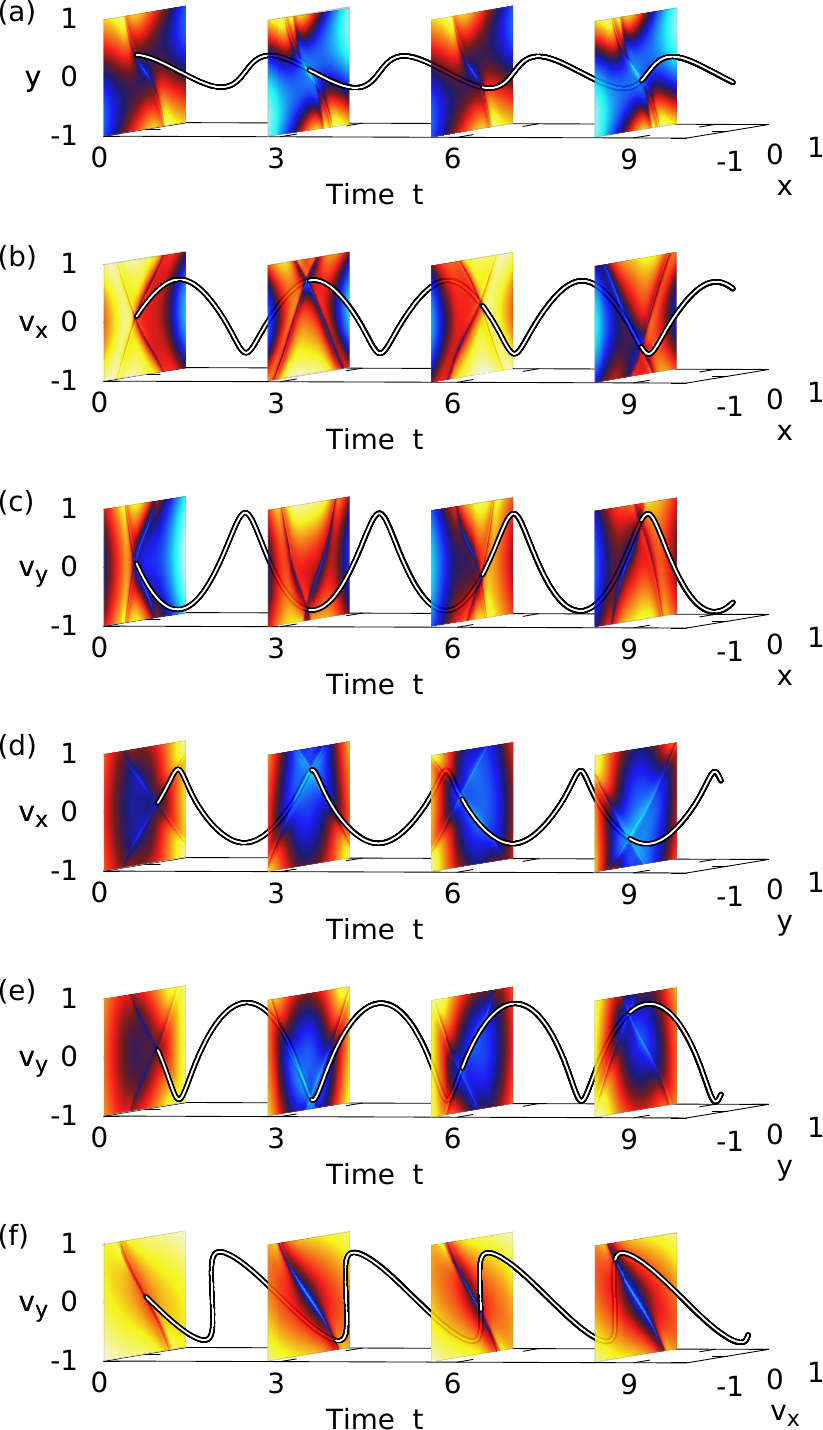}
\caption{%
Complete time-dependent phase space representation of a periodic orbit (solid 
line) and the corresponding LD~\eqref{eq:LD}.
The comparison shows that the PO's coordinates are equivalent to the LD 
minimum throughout.
The panels (a)--(f) show the time-evolution of both these objects for the 
different two-variable combinations of the variables $x,y,v_x,v_y$.
}
\label{fig:PODS-phase-space}
\end{figure}
%%%%%%%%%%%%%%%%%%%%%%%%%%%%%%%%%%%%%%%%%%%%%%%%%%%%%%%%%%%%%%%%%%%%%%%%%%%%%%%

A complete phase space visualization of the 
dynamics and LD of a single PO is shown in 
\FIG~\ref{fig:PODS-phase-space}.
Each panel shows the projection of the trajectory onto two of the 
four phase space variables.
The trajectory is shown as a solid curve.
A series of four colored contour planes of the LD is shown at selected
times.

The PO intersects with all phase space cuts exactly where the 
LD minimum is located, once again verifying the relation in 
\EQ~\eqref{eq:LD-PO}.
Thus, the dynamical picture in \FIG~\ref{fig:PODS-phase-space} 
illustrates that 
this relation is valid in the whole phase space and for any time $t$.

\subsection{Determining instanton trajectories in 2D}
\label{sec:results-tunneling}

We now demonstrate the use of the LD-PO method for
constructing instanton trajectories through a barrier.
The instanton trajectory is a PO in the inverted potential $-V$ and, in a path 
integral formalism, it is directly related to the tunneling of a particle 
through the barrier.\cite{mill74b,mill75b,kleinert2009path,stoof97,Wunner08a} 

We use the potential
\begin{align}
\begin{split}
 V(x,y) = &
 \frac{1}{2x^2} + \frac{1}{8y^2} + 2\gamma_x^2x^2 + 2\gamma_y^2 y^2 
 + \frac{a}{2\sqrt{2}x^2y} \\
 &+\frac{1+\kappa^2-(3x^2\arctan\sqrt{\eta})/(2y^2\sqrt{\eta})}
        {12\sqrt{2\pi}x^2y\eta} \,,
\end{split}
\label{eq:dip-BEC-H}
\end{align}
where $\kappa=(x/y)^2$, $\eta=\kappa/2-1$, and $\gamma_{x,y}, a$ are free 
parameters in arbitrary units.
This potential describes within a mean-field approximation and a Gaussian 
approach to the wave function a dipolar Bose-Einstein condensate (BEC) in an 
axisymmetric, harmonic trap, 
where the variables $x,y$ can be interpreted as the mean extensions of the wave 
function.\cite{Huepe1999,Huepe2003}
The parameters $\gamma_{x,y}$ are the strength of the external traps (which we 
define by $\gamma_x^2\gamma_y=34\,000$ and $\gamma_y/\gamma_x=6$) and $a=0.1$ 
is the s-wave scattering length describing the contact interaction of two 
bosons.
The physical meaning of \EQ~\eqref{eq:dip-BEC-H} is as follows:
The potential exhibits a local minimum at $x\approx0.013$ and $y\approx0.023$ 
which corresponds to the metastable ground state of the BEC and it diverges to 
$V\to-\infty$ for $x\to0$.
Both these regions are separated by a barrier and its crossing physically means 
the collapse of the BEC.
There are two different ways to reach the collapsed state:
One is the classical crossing of the barrier at high energies which is 
referred to as the thermally induced coherent collapse of the 
condensate.\cite{Huepe1999,Huepe2003,junginger2012d,Junginger2013b}
The other possibility is the tunneling through the barrier in the limit of zero 
temperature referred to as 
macroscopic quantum tunneling.\cite{stoof97,Wunner08a}
As mentioned above, the theoretical description of the tunneling through the
barrier in a path integral formalism \cite{kleinert2009path} is directly related 
to the properties of 
the instanton trajectory which is the corresponding PO at the saddle of the 
inverted potential $-V$.
In the following, we therefore apply the method described in 
\SEC~\ref{sec:theory} to those trajectories.

%%%%%%%%%%%%%%%%%%%%%%%%%%%%%%%%%%%%%%%%%%%%%%%%%%%%%%%%%%%%%%%%%%%%%%%%%%%%%%%
\begin{figure}[t]
\centering
\includegraphics[width=0.9\columnwidth]{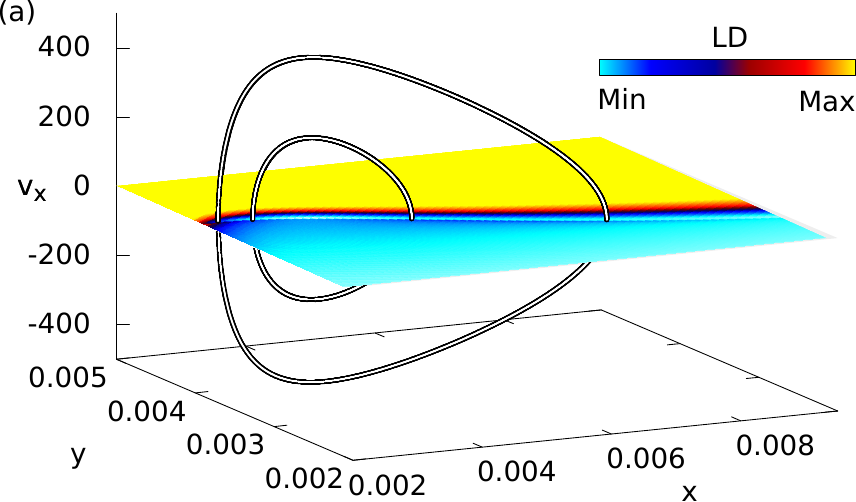}
\includegraphics[width=0.9\columnwidth]{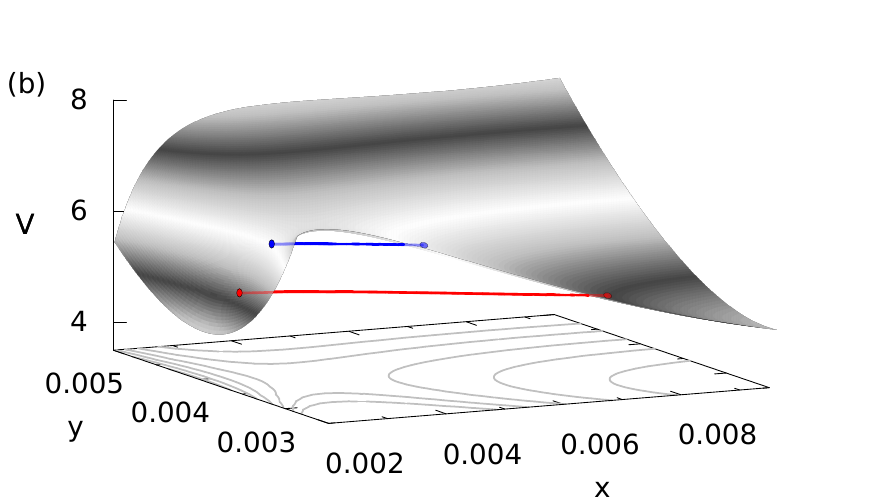}
\caption{%
Visualization of two POs (solid lines) at different energies below the 
activation energy on the inverted potential $-V$ of 
\EQ~\eqref{eq:dip-BEC-H}:
(a) 
A colored contour surface of LD values obtained at $\tau=10$ 
with POs cutting through the minimum values.
(b) 
Configuration space projection of the instanton 
trajectories draw on the potential.
}
\label{fig:tunneling}
\end{figure}
%%%%%%%%%%%%%%%%%%%%%%%%%%%%%%%%%%%%%%%%%%%%%%%%%%%%%%%%%%%%%%%%%%%%%%%%%%%%%%%

Figure~\ref{fig:tunneling}(a) presents an LD cut through the $x$-$y$-plane 
calculated for the inverted potential, $-V$ in \EQ~\eqref{eq:dip-BEC-H}, and it 
can be seen that this LD portrait also exhibits a minimum valley.
In analogy to the case in \SEC~\ref{sec:results-classical},
the corresponding phase space coordinates are initial values of 
classical POs in the inverted potential
as indicated by the solid lines.
The phase space coordinates at the minimum valley again correspond to initial 
points of the corresponding POs.
We note here that, because of the divergence of the potential 
\eqref{eq:dip-BEC-H} for $x\to0$, we have cut off the trajectories for the LD 
calculations when they have reached $x\leq 0.001$. 
This prevents one from facing numerical problems with the diverging potential 
for $x\to0$, but it does not affect the search for the POs because they never 
reach this region.

Since the POs in \FIG~\ref{fig:tunneling}(a) have been calculated for the 
inverted potential $-V$, the energy of the particles is below the saddle point 
energy for the real potential $+V$.
In \FIG~\ref{fig:tunneling}(b), we present the configuration space projection 
of the respective POs in the original potential $+V$, where the trajectories 
open up the tunnel connecting the two regions on both sides of the barrier.

\section{Conclusion and outlook}

From the relation of the PO being located at the intersection of the stable and 
unstable manifolds together with the fact that the LD exhibits extremal, 
\ie~minimal, values on these manifolds, we have presented a general 
construction scheme for this type of periodic orbits.
Our LD-PO method is solely based on a minimization of the LD in 
phase space (or a subspace of constant energy, respectively).
Neither boundary conditions nor any previous 
knowledge (\eg~the period $T$) are required.
The procedure is easy to implement formally,
as well as numerically, because 
the LD can be obtained on the fly when the trajectory is integrated and the 
minimization can be performed \eg~using standard 
algorithms.\cite{NumericalRecipes2007}
Since the LD-PO method is based on minimizing a scalar phase space function, 
it can also be applied directly in higher-dimensional systems.

We have shown the applicability of the procedure to both the cases of energies 
above the saddle point, \ie~classical POs, as well as energies below the saddle 
point where the solutions represent the instanton trajectory related to the 
tunneling through the barrier.
In both cases, we have identified minimum valleys of the LD in phase space 
which provide initial values of the POs at the respective energies.

We further observe that not only 1-cycles,
but also $n$-cycles, can be obtained by the LD-PO method.
Such POs of higher order emerge as local minima of the LD.
They can be distinguished from each other by the LD's actual value which is 
a measure of the PO's cycle order.
Possible applications lie in the general 
construction of POs in various systems and the extraction of periodic orbit 
dividing surfaces or instanton trajectories in reaction dynamics.

\begin{acknowledgments}
AJ acknowledges the Alexander von Humboldt Foundation, Germany, 
for support through a Feodor Lynen Fellowship.
RH's contribution to this work was supported by
the National Science Foundation (NSF) through Grant
No.~CHE-1700749.
\EDITS{This collaboration has also benefited from support by the
people mobility programs of the European Union,
and most recently,
the Horizon 2020 research and innovation programme under grant agreement 
No.~734557.}
\end{acknowledgments}
%\vfill ~

\bibliography{q04paper}
\end{document}